\documentclass[10pt,showpacs,preprintnumbers,amsmath,amssymb]{revtex4}

\usepackage{epsf}
\usepackage{graphicx}  
\usepackage{dcolumn}   
\usepackage{bm}        

\newcommand{\be}{\begin{equation}}
\newcommand{\en}{\end{equation}}
\newcommand{\ba}{\begin{array}}
\newcommand{\ea}{\end{array}}

\begin{document}

\preprint{ITP-CAS/3-2005}

\title{Friedmann cosmology on codimension 2 brane with time dependent tension}

\author{Hongsheng Zhang\footnote{E-mail: zhanghs@itp.ac.cn}
 ,~Qi Guo\footnote{E-mail: Guoqi@itp.ac.cn}
  and Rong-Gen Cai\footnote{E-mail: cairg@itp.ac.cn}}
\affiliation{Institute of theoretical physics, Chinese academy of
sciences, Beijing P.O. Box 2735, China}
\date{ \today}

\begin{abstract}
A solution of codimension 2 brane is found for which 4 dimensional
Friedmann cosmology is recovered on the brane with time dependent
tension, in the Einstein frame. The effective parameter $p/\rho$
of equation of state on the brane can be quintessence like, de
Sitter like or phantom like, depending on integration constants of
the solution.

\end{abstract}


 \maketitle
 In 1983 Rubakov and  Shaposhnikov  presented an
 extraordinary
 model \cite{rubakov} to solve the cosmological constant problem,
 in  which the cosmological constant appeared as an integration
 constant. In this model the effective cosmological constant
was  free of the zero point quantum energy. The zero
 point energy only warps the extra dimension, and at the same time our
 4 large dimensional spacetime keeps flat. This is similar to the case
 that although in realistic
 universe the 4 dimensional curvature may be high but the 3 dimensional
 spatial curvature is almost zero.

  During recent years codimension 1 brane model has been
 fairly well understood. Also the cosmology in frame of codimension 1
 brane  has been thoroughly studied, for a review see \cite{lang}.
 Generally speaking codimension 1 cosmology of RS model implies Friedmann
 equation like
 \be
 H^2\sim \rho+\frac{\rho^2}{\sigma},
 \en
where $\sigma$ is the brane tension, $\rho$ is the energy density
of matter on the brane. We find the standard cosmology recovers
when the energy scale is low enough and the corrected Friedmann
equation gives notable effect when the energy scale is high enough
. But to the codimension 2 case, we get into a much more weird
situation: although the effective cosmological constant is
independent of the tension of the brane, to our surprise, we
appear to be very restricted in our permitted brane
energy-momentum. Typically, a brane in its ground state has a very
special energy-momentum tensor, which is isotropic and has the
property that energy density=--pressure \cite{navarro}, that is to
say, the only spacetime allowed is maximally symmetric space, no
Friedmann cosmology with generic parameter of state equation
$p/\rho$ can be set up on the brane.

 To circumvent this embarrassed situation and get a
viable cosmology on codimension 2 brane, different authors  have
put much effort. Clearly, two possible ways are available. First,
inspired by the early work on cosmic string \cite{gott}, a fat
brane is a natural way to circumvent this problem. A such model
has been considered in \cite{vinet}, in which  a codimension 2
braneworld with spherical extra dimensions compactified by
magnetic flux is studied. Assuming Einstein gravity, it shows that
when the brane contains matter with an arbitrary parameter of
equation of state, and find that the universe expands consistently
with standard Friedmann cosmology. However, the cost is the
relation between the brane tension and the bulk deficit angle
becomes ~$\delta \sim \rho-3p$~ for a general equation of state.
This relation does not imply a self-tuning of the effective 4D
cosmological constant. So this progress is bankrupt in the merit
that the effective cosmological constant is independent of the
brane tension, which is not what we want. In \cite{navarro} the
action includes classical Hilbert action, we may curious whether
or not a higher order Ricci scalar terms can circumvent the
restriction energy density=--pressure on the brane. That is the
second accessible possibility. A proposal along this direction has
been presented in \cite{bost}, in which a general braneworld in 6
dimensional Einstein-Gauss-Bonnet gravity is studied. It shows how
the 4 dimensional Einstein equation is recovered for the induced
metric and matter on the brane. It also shows that relaxing
regularity of the curvature in the vicinity of the brane gives
rise to an additional possible correction to the Einstein
equation. Now within this proposal the 4 dimensional Einstein
equation recovers on the brane, therefore the Friedmann cosmology
can be obtained on the brane without question. But we may bring
forward the following questions to this suggestion: why in pure
classical domain we must consider higher order correction to
revive Einstein equation on the brane? If the $R^2$ term
correction gives the answer that we need, as the authors stressed,
since $O(R^2)$ corrections to the Hilbert Lagrangian do arise in
the low energy limit of string theory, the inclusion of this type
of term could be regarded as mandatory if one wants to embed any
braneworld solution into string/M-theory, how about the more
higher order corrections? They may give ``false'' results.

 In this paper we successfully recover almost realistic Friedmann
 cosmology on the
 brane in frame 6 of dimensional Einstein gravity with a 3 brane on
 which the
 tension is variable. If the brane tension originates from the
 quantum zero energy, it would decay with the expansion of the
 universe, as suggested in
 \cite{schu}. That article addresses the question
of whether non-perturbative effects of self-interacting quantum
fields in curved space-times may yield a significant contribution
to the tension. Focusing on the trace anomaly of quantum
chromo-dynamics (QCD), a preliminary estimate of the expected
order of magnitude yields
 \be
 \rho=n H,
 \label{qcd}
 \en
where $n=m^3$, $m$ is  energy scale of QCD chiral phase
transition, $\rho$ is vacuum energy (brane tension) and $H$ is
Hubble parameter. $H$ is defined as
 \be
 H=\frac{1}{a} \frac{da}{dt}.
 \en

 As a natural generalization of $Z_2$ symmetry in codimension 1
 brane model, the metric ansatz is taken to be
  \be
  ds^2=-dt^2+a(t)^2d\vec{x} ^2+dr^2+b(t)^2 v(r)^2 d\theta^2,
  \label{metric}
  \en
  where $d\vec{x}^2$ is the metric of 3 dimensional maximally
  symmetric space, $v(r)=\sin(r),~~r,$~~or,~~ $\sinh(r)$,
  depending on the internal space is a sphere, plane or
  pseudosphere. To spherical case the internal space is  compact
  by nature.   And to case of plane or
   pseudosphere the internal space is compactfied by proper identification,
   that is to say, we identify all the points $r=$constant.  The brane stands at
  $r=0$. Supposing the metric is regular on the brane, it is easy
  to get the familiar result in codimension 2 infinite thin braneworld models
  \cite{navarro, bost}
 \be
  1-b=4 G_6 \sigma,
  \en
 where $\sigma$ is the brane tension which  equals the energy
 density of the brane,  and $G_6$ denotes the 6 dimensional Newton constant.
  According to Schutzhold's investigation the brane
 tension is proportional to the Hubble parameter, as shown in
 equation (\ref{qcd}). In its original form the coefficient is
 related to the energy scale of QCD chiral phase transition.
  Although it gives
 right energy scale of our observed cosmological constant, we
 should not restrict us to the energy scale of QCD
 phase transition from the theoretical view because various phase
  transitions may happen in the history of our
 universe.  Here we
 adopt a more generic case,
 \be
 \rho(t)=\sigma(t)=\frac{1}{4G_6} c H,
 \label{rho(t)}
 \en

 in which $c$ is a constant, it can contain all the contributions of
 phase transitions ever happened on the brane. Note that
 the dimension of $c$ is $[l]$. Thus a smaller $c$ corresponding
 to a higher energy scale. Einstein field equations of this
 system is
  \be
  ^{(6)}G_{AB}=8\pi G_6 T_{AB},
  \en
where
  \be
  T_{AB}=T_{AB}^{(bu)}+T_{AB}^{(br)},
  \en
 $^{(6)}G_{AB}$ denotes 6 dimensional Einstein tensor, $T_{AB}$
 is the total energy momentum tensor, $T_{AB}^{(bu)}$ and $T_{AB}^{(br)}$
 represent the energy on the brane and in the bulk respectively,
 and $G_6$ is the 6 dimensional Newton constant. The brane energy
 momentum is shown clearly in matrix form,
  \be
 T_{AB}^{(br)}=\left(
  \ba{ccc}
  \hat{T}_{\mu\nu}^{(br)} \delta(r) &~&~\\
  ~&0 &~\\
  ~ &~ &0
  \ea
  \right),
 \en
 where $\hat{T}_{\mu\nu}^{(br)}=-\sigma (t) g_{\mu\nu}$, and $\mu$ or
 $\nu$ runs from 0 to 3.
 We do not try to specify $T_{AB}^{(bu)}$ here because we will find later that
 it has to take a form of non perfect fluid for a self consistent
 solution under condition (\ref{rho(t)}).
  Now consider the functions $a(t)$ and $b(t)$ in the following
  forms
  \be
  a(t)=c_3 e^{\frac{2(c_2+t)}{c}} \cosh^2[\frac{(t+c_2)\sqrt{4+2 c^2
  c_1}} {2c}],
  \label{solution}
 \en
 and
  \be
  b(t)=-1-\sqrt{4+2 c_1 c^2} \tanh[\frac{(t+c_2)\sqrt{4+2 c^2
  c_1}} {2c}].
  \label{angle}
  \en
  Here $c_1,~c_2,~c_3$ are 3 integration constants. We shall explain
  their physical meanings later in Einstein frame.  One can verify
  that $a(t)$ and $b(t)$ solve the 6 dimensional Einstein equation with
  non perfect fluid in the bulk. We can prove it is the unique
  solution satisfying the following conditions:\\
  1. The metric is spatially flat, that is,  $d\vec{x}^2$ in (\ref{metric}) is
  Euclidean.\\
  2. Although we permit the matter in the bulk to be non perfect fluid,
  we impose the condition
   $-{T^{(bu)}}^0_0={T^{(bu)}}^1_1={T^{(bu)}}^2_2={T^{(bu)}}^3_3$,
   to ensure
  that it behaves as perfect fluid along the brane. From the
  physical viewpoint it is a reasonable requirement.\\
    Here we will present some other notes on this solution.
 Both $a(t)$ and $b(t)$ are internal--curvature--independent, that
 is, they do not depend on the Gaussian curvature of the
 internal space, which means, spherical, flat and pseudospherical internal
 spaces correspond to the same $a(t)$ in (\ref{solution}).
 The tension of the brane is time dependent, which most
 probably leads to the time dependence of the energy momentum
 tensor in the bulk. We find it is just the case. Explicitly,
  \be
 -{T^{(bu)}}^0_0=\frac{ 3\cosh[f]^{-3} \left( (4-c_1
 c^2)\cosh[f] +(12+5c_1 c^2)\cosh[f]+2l(4+(4+\frac{l^2}{2}
  \cosh[f])\sinh[f]) \right)}{16c^2 \pi G_6(1+l\tanh[f])},
  \en
 and
 \be
 {T^{(bu)}}^4_4={T^{(bu)}}^1_1+\frac{1}{16\pi G_6} \left( -\frac{3}{2} l^2
 \cosh[f]^{-2} +6(4+c_1 c^2+2l\tanh[f])\right),
 \en

 where $f=\frac{l(c_2+t)}{2c}$ and $l=\sqrt{4+2 c_1 c^2}$.
  $-{T^{(bu)}}^0_0={T^{(bu)}}^1_1={T^{(bu)}}^2_2={T^{(bu)}}^3_3$
  and ${T^{(bu)}}^4_4$ are also
  internal--curvature--independent. The situation is different to
  ${T^{(bu)}}^5_5$, it is internal--curvature--dependent. To avoid
  complexity we only give specific form of the flat internal space
  case,
  \be
  {T^{(bu)}}^5_5={T^{(bu)}}^1_1+\frac{72+30c_1 c^2-\frac{33l^2}{2} \cosh[f]^{-2}+6l
  +\left(6+c_1c^2-\frac{l^2}{2}\cosh[f]^{-2} \right) \tanh[f]} {8\pi
  G_6c^2(1+l\tanh[f])}.
  \en
  To the cases of sphere and pseudosphere the discussions are
  similar. This model also possesses another remarkable property.
  There is a non zero energy flux along the $r$ direction
  emerging in the bulk. In the case of flat internal space, it has
  the  form as
  \be
  w_4=-{T^{(bu)}}_{40}=\frac{l^2 \cosh[f]^{-2}}{16\pi G_6 cr(1+l\tanh[f])}.
  \label{flux}
  \en
 This is exactly a rational result. We see the vacuum energy of the
 brane is decaying, but there is nothing  on the brane to absorb
 the energy decayed \footnote{Recently a model of Friedmann
  cosmology with decaying vacuum density in 4 dimensional frame
 has  been discussed by H. A. Borges and S. Carneiro in
 gr-qc/0503037, in which a decaying vacuum term
leads to matter production.}. Therefore the only way out of this
energy  is to
 flow into the bulk. There is a singularity in (\ref{flux}) at
 $r=0$, where the brane habitats. Any finite energy flux flowing from the
 brane must be divergent  at the neighborhood of the
 brane , since the brane is infinite thin.
 In
 respect to these results we are able to predict the energy flux
 vanishes at $r=\pi$ in case of spherical internal space and
 slopes
 more sharply with respect to $r$ in case of pseudospherical internal
 space. In fact these are just the main differences of bulk energy
 fluxes  between flat
 and spherical or pseudospherical internal spaces. Still another
 point deserves to address. There is a deficit angle in the extra dimension
 due to the existence of the brane in all previous works on the codimension 2 brane
 models \cite{bost,vinet,navarro}. Here, however, from (\ref{angle})
 it is clear the ``deficit angle" is negative, or, there is a surplus
 angle $2\pi(|b|-1)$ since $|b|>1$. For example the internal space
 becomes a mandarin orange rather than an American football in case of spherical
 internal space.

  If we believe only Einstein frame makes sense in our real world,
  it is not obvious whether or not the solution (\ref{solution})
  really represents a generic Friedmann universe. Now we turn to
  Einstein frame, in which the metric becomes
  \be
  ds_E^2=-|b|dt^2+|b|a^2d\vec{x}^2.
  \en
 Define proper time
  \be
  d\tau=|b|^{1/2}dt,
  \en
  and the proper scale factor,
  \be
  \tilde{a}=|b|^{1/2} a.
  \en
  Define cosmic
 velocity
  \be
  V_e=\frac{d\tilde{a}}{d\tau},
  \en
 and cosmic acceleration
  \be
  A_c=\frac{d^2\tilde{a}}{d\tau^2}.
 \en
 Now it is time for explaining the physical meanings of
 the constants $c_1,~ c_2,~ c_3$.
  Obviously $c_3$ is only a constant scale factor, and $c_2$
  represents a relocation of initial time in (\ref{solution}). To
  make out the meaning of $c_1$, we calculate the proper Hubble
  parameter
  \be
  \tilde{H}=\frac{V_e}{\tilde{a}}.
  \en
  We find
  \be
  \tilde{H}(t=-c_2)=\frac{3}{c}+\frac{c_1 c}{2}.
  \en
  It is clear that in some sense $c_1$ indicates initial cosmic velocity.
  Naively when
  $c^2c_1=-2$ this solution degenerates to a de Sitter brane
  without angle deficit in the bulk in view of $b=-1$. Even this
  most simple case is not the trivial one studied before, because in our case
  $\sigma$ is not a constant all the same, while
  $\sigma=$constant in all present studies.

  The most significant parameters from the viewpoint of
  observations are parameter of state equation $w=\frac{\rm pressure}
  { \rm energy~density}$ and the deceleration parameter $q$ of the total cosmic fluid. The two
  parameters are
  closely related to each other.
   Therefore, we calculate the effective $w$ and $q$ corresponding
 to the proper scale factor and proper time.
  By definition
 \be
 w_{\rm eff}=\frac{p_{\rm eff}}{\rho_{\rm eff}}=-\frac{V_e^2+2aA_c}{3V_e^2},
 \en
 and
  \be
  q=-\frac{aA_c}{V_e^2},
  \en
  we arrive at
   \be
  q=\frac{4
  c_1c^2(3+c_1c^2)\cosh[f]-(10+\frac{5}{2}l^2+l^4)\cosh[f]+
  3(4+3c_1c^2+2c_1c^2l\sinh[f]-2(3+c_1c^2)l\sinh[f])}{
  (c_1-2(3+c_1c^2)\cosh[f]-3l\sinh[f])^2u},
 \label{dc}
  \en
  and
  \be
  w_{\rm eff}=-\frac{1}{3}+\frac{2}{3}q,
  \label{weff}
  \en

 where $u=\sqrt{1+l\tanh[f]}$. The equations are rather
 complicate. It needs some labour to check when $c^2c_1>-2$,
 the universe finally becomes quintessence like phase, which
 in some sense can
 simulate dark matter plus cosmological constant universe; when
 $c^2c_1=-2$, it is a de Sitter phase; and when $c^2c_1<-2$,
 the universe goes into a phantom like phase.   This is
 our main result: Friedmann universe of rather arbitrary
 parameters of state equations can recover on the
 codimension 2 varying--tension--brane. Different types of the
 universe depend on different choices of the integration constants. We
 show this result clearly in some figures. The figures illustrate that the
 evolution of the universe only depends on $c^2c_1$. Recall equation
 (\ref{rho(t)}), $c$ is the vacuum energy on the brane. It is
 clear we can always choose a proper $c_1$ to counter the effect
  of an arbitrarily
 large $c$. Accordingly the most important virtue of prior models with 2
 extra dimensions \cite{rubakov,navarro}
 is retained in our model. One may be afraid that if
 $c^2c_1<-2$, the arguments of the hyperbolic functions in
 (\ref{dc}) become imaginary numbers. But we find the final
 results always keep real even if the imaginary arguments appear.

 \begin{figure}
\centering
 \includegraphics[totalheight=2in]{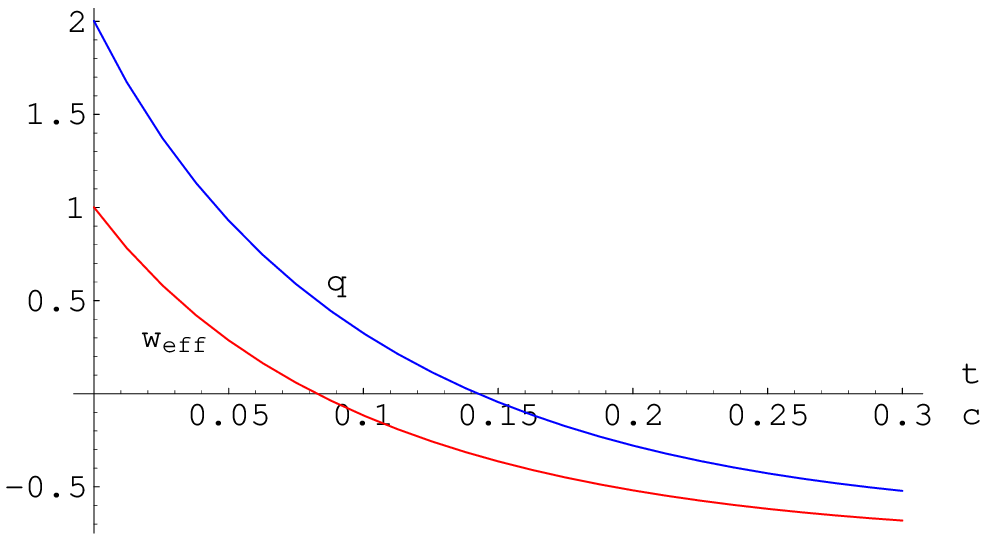}
 \caption{$\frac{t}{c}$ versus $w_{\rm eff}$  and $q$: the left part.\\
  To  showing clearly the properties of $w_{\rm eff}$  and $q$ we
  plot two figures with the same set of parameters, in which $t$
  is in different intervals of real line. To this figure
  $\frac{t}{c} \in(0, 0.3)$, and to the next $\frac{t}{c}\in (0.3, 3)$.
   In this figure $c_1c^2=1$ , $c_2$ has been tuned such that
   $w_{ \rm eff}=1$ when $t=0$.}
 \label{quintde}
\end{figure}

\begin{figure}
\centering
 \includegraphics[totalheight=2in]{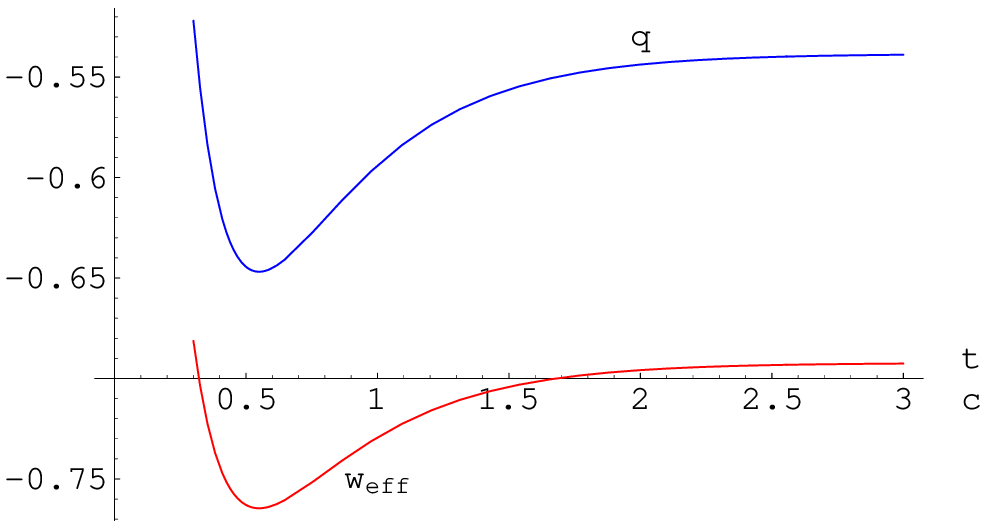}
 \caption{$\frac{t}{c}$ versus $w_{\rm eff}$  and $q$: the right part.
 \\ ~~~In this figure the parameters are chosen
 as  same as the last figure.  These two figures
 show that a universe undergoes stiff matter, radiation, dust
 like phases and finally becomes a quintessence like universe.
 From the observations we know that $w_{\rm eff}$ is declining,
  so we may think that the current universe
 corresponds to the coordinate $\frac{t}{c}=0.3$. If it is true in the future
  $w$ will decrease for some time and then increase, eventually reach
  approximately --0.68. \\
   Because the age of our universe is model dependent we can not
    obtain zero energy from $\frac{t}{c}=0.3$ by  using the value
   of $t$ in the standard model. But we point out that
   for any given $c$ we can always obtain such a universe,
   as shown in figure 1 and figure 2,
   by tuning $c_1$. }
 \label{quintq}
\end{figure}

 \begin{figure}
\centering
 \includegraphics[totalheight=2in]{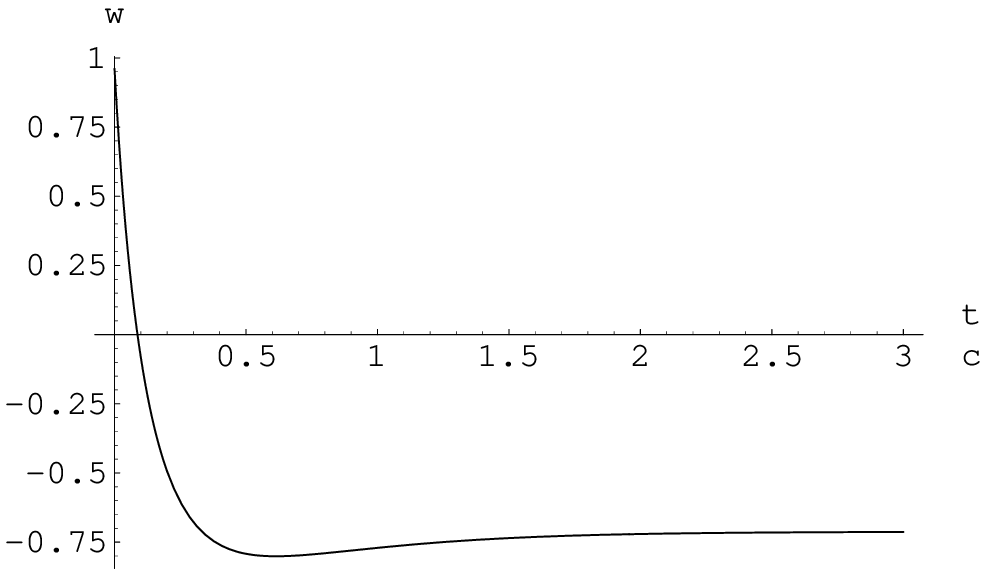}
 \caption{$\frac{t}{c}$ versus $w$.\\ In this figure $c_1c^2=-0.5$. This
 figure shows that we can adopt parameters to ensure the final
 value of $w$ equals 0.72, which may offer some implications on the
 coincidence problem.}
 \label{coin}
\end{figure}

\begin{figure}
\centering
 \includegraphics[totalheight=2in]{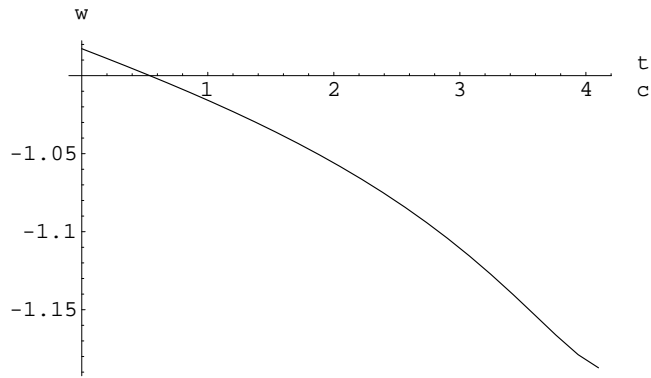}
 \caption{$\frac{t}{c}$ versus $w$. \\
 In this figure $c_1c^2=-2.1$. A phantom
 dominated universe emerges under this parametrization.}
 \label{phan}
\end{figure}

 To sum up: in this paper we discuss a codimension 2 braneworld model, in which
 the brane tension depends on time. We find a unique solution
 under some reasonable considerations. According to this solution
 Friedmann cosmologies , which permit different parameters of state
 equations, can be set up on the brane in Einstein frame.
 This is sharply different
 from the previous researches on this topic. At the same time our
 model keeps the most impressive character of the codimension 2 brane
 model---the effective cosmological constant is free of the
 absolute value of the brane tension. In our model the parameter
 of state equation only depends on the integration constants, that
 is to say, it circumvents the famous $10^{120}$ contradiction. So
 the problem becomes why  the integration constants are taking those
 values.

 We will also point out some questions to be discussed future in our
 model. Follow from figure \ref{quintde}, if it is believed that we are
 living at $t/c=0.3$, there exists another coincidence problem, which is
 essentially the choices of integration parameters, all the same. The
 asymptotic behavior of the universe in figure \ref{coin} overcomes the
 coincidence problem, but it is difficult to simulate the history
 of our universe, especially the history of structure formation.
 It is still an open problem in present stage.


{\bf Acknowledgments:} This work was supported in part by a grant
from Chinese academy of sciences, a grant No. 10325525 from NSFC,
and by the ministry of science and technology of China under grant
No. TG1999075401.


\begin{thebibliography}{99}


\bibitem{rubakov}
V. A. Rubakov and M. E. Shaposhnikov,
 Phys. Lett. B 125, 139 (1983).


\bibitem{lang}
D. Langlois, hep-th/0209261.


\bibitem{navarro}
 I. Navarro,  JCAP, 0309 (2003), 004, hep-th/0302129;

  S. M. Carroll and M. M. Guica, hep-th/0302067.

\bibitem{gott}
J. R. Gott III, APJ, 288 (1985), 422.

\bibitem{vinet}
 J.  Vinet and J. M. Cline, Phys.Rev. D70 (2004) 083514, hep-th/0406141.
\bibitem{schwi}
J. Schwindt and C. Wetterich, hep-th/0501049.

\bibitem{bost}
P. Bostock, R. Gregory, I. Navarro and J. Santiago, Phys.Rev.Lett.
92 (2004), 221601, hep-th/0311074.

\bibitem{schu}
R. Schutzhold, Phys. Rev. Lett. 89, 081302 (2002), gr-qc/0204018.



\end{thebibliography}
\end{document}